\documentclass[review,sort&compress,number]{elsarticle}

\usepackage[T1]{fontenc}
\usepackage[utf8]{inputenc}
\usepackage{lmodern}
\usepackage{microtype}
\usepackage[english]{babel}
\usepackage{enumitem}
\usepackage{amsmath}
\usepackage{amssymb}
\usepackage{graphicx}
\usepackage{wrapfig}
\usepackage{float}
\usepackage{xcolor}
\usepackage{algorithm}
\usepackage{algorithmic}
\usepackage{multirow}
\usepackage{bbm}
\usepackage{gensymb}
\usepackage{textcomp}
\usepackage{gensymb}
\usepackage{lineno}

\usepackage{fancyhdr}
\begin{document}

\begin{frontmatter}
\title{Multifractal analysis of the time series of daily means of wind speed in complex regions}
\author{Mohamed Laib$^1$, Jean Golay$^1$, Luciano Telesca$^2$ and  Mikhail Kanevski$^1$}
\address{$^1$IDYST, Faculty of Geosciences and Environment, University of Lausanne, Switzerland. \\
$^2$CNR, Istituto di Metodologie per l’Analisi Ambientale, Tito (PZ), Italy \\
Corresponding author: mohamed.laib@unil.ch}

\begin{abstract}

In this paper, we applied the multifractal detrended fluctuation analysis to the daily means of wind speed measured by 119 weather stations distributed over the territory of Switzerland. The analysis was focused on the inner time fluctuations of wind speed, which could be more linked with the local conditions of the highly varying topography of Switzerland. Our findings point out to a persistent behaviour of all the measured wind speed series (indicated by a Hurst exponent significantly larger than 0.5), and to a high multifractality degree indicating a relative dominance of the large fluctuations in the dynamics of wind speed, especially in the Swiss plateau, which is comprised between the Jura and Alp mountain ranges. The study represents a contribution to the understanding of the dynamical mechanisms of wind speed variability in mountainous regions. 

\end{abstract}


\end{frontmatter}

%\linenumbers
\section{Introduction}
\label{intro}
Wind has been gaining an increasing attention in the context of renewable energy because it represents one important substituent to conventional fuels that should play a major role in the future energy mix \cite{Edenhofer2012}. In fact, wind power has the advantage to be widely produced (in the three consecutive years, from 2013 to 2015, the world total wind capacity grew up more than $10\%$ \cite{Dai2017}) with little environmental pollution, becoming economically competitive as a type of clean energy source capable of withstanding environmental damage and avoid future crises \cite{Nematollahi2016}.

In mountainous regions like the Alps, wind speed highly changes not only in time but also in space. Topography, in fact, influences strongly the wind speed \cite{Barry1992, Mortensen1998}. Ridge crests, deep valleys, or other irregular landscapes are important orographic features that can exert an influence on boundary layer flows \cite{Etienne2010}. The Alps are characterized by many local climatic phenomena, natural channelling effects, and thermally induced circulations that make, for instance, wind speed very high at one location but very slow in a near valley, revealing a large variability and discontinuous character within small areas, and making the spatial interpolation of wind speed quite arduous   \cite{Tveito2008}. The analysis of wind speed within the atmospheric boundary layer is always challenging, since it represents a largely fluctuating and non-linear component of atmospheric flows, whose space-time variability can be high \cite{Etienne2010}. In addition, the modelling of wind speed and extreme events, as well as the regionalisation schemes of wind speed and direction is rather difficult \cite{Goyette2001}, due to the presence of many turbulence effects and roughness factors \cite{Stull1988}. However, some methods have been developed to overcome the problems related to the regionalisation of wind speeds \cite{Porch1987, Palomino1995, Nielsen1999, Schaffner2006, Etienne2010}, by using correction factors related to topography (slope angle, altitude, land-form characteristics) that, added to the calculation of wind speed, have enabled a better adjustment of the results with the observations.

Several studies have been performed on the wind speed field over the territory of Switzerland \cite{Jungo2002, Schaffner2006}, which is characterized by a topographically complex terrain. Etienne et al.  \cite{Etienne2010} applied the Generalized Additive Models (GAMs) to regionalise wind speeds measured at the Swiss weather stations by means of a number of physiographic parameters. They succeeded in providing reliable wind predictions on the basis of the $98^{th}$ percentile of the daily maximum wind speed, and found a dependence of wind speed upon the altitude and roughness of the mountain shapes. Jungo et al. \cite{Jungo2002} applied the Principal Component Analysis (PCA) and the Cluster Analysis (CA) to several Swiss meteorological stations. They clustered these structures based on their daily gust factors, depending on the weather type. As a result, the obtained clusters of stations, whose spatial distribution depended on the complexity of terrain, exhibited a comparable variability to the daily gust factor and to their response to the weather forcing. Weber and Furger  \cite{Weber2001} applied an automated classification scheme to one year wind data and found $16$ distinct near-surface wind flow patterns, whose knowledge is important because they can form intricate patterns such as large-scale winds and locally forced wind systems interplay. Robert et al. \cite{Robert} applied general regression neural networks (GRNN) as a non-linear regression method to interpolate monthly wind speeds in complex Alpine orography, using as training data those coming from Swiss meteorological networks to get the relationships between topographic features and wind speed.

In this study, we aim at analysing the time series of daily mean of wind speed recorded by a wide monitoring network covering all the territory of Switzerland. Because of the complexity of the terrain, related to the large variability of topographic conditions that characterize Switzerland, wind speed is featured by a complex time dynamics. In order to investigate the dynamical properties of wind speed time series, we use
the multifractal detrended fluctuation analysis (MFDFA) to identify correlations, persistence, intermittency, and heterogeneity.

Since the investigation of multifractal behaviour in wind speed has been becoming an important topic only recently, just a few studies have been carried out so far. Kavasseri and Nagarajan \cite{Kavasseri2005} showed that the multifractality found in four time series of hourly means of wind speed in USA could be explained by  fitting the data with a binomial cascade multiplicative model. Telesca and Lovallo \cite{telesca2011} found that most of the multifractality of the wind speed series recorded at several heights from the ground from $50m$ to $213m$ was due to the different long-range correlations in small and large speed fluctuations. Fortuna et al. \cite{Fortuna2014} applied the MFDFA to several hourly wind speed series in Italy and USA and found that the multifractal width ranged in a quite close interval of values between $0.39$ and $0.59$.  De Figueiredo et al.  \cite{DEFIGUEIRA2014} found that the mean and the maximum of four wind speed time series in Brazil were all persistent, but the maximum was more multifractal than  the mean. Piacquadio and De La Barra  \cite{PIACQUADIO2014} suggested the use of some key multifractal parameters of wind speed as local indicator of climate change. Telesca et al. \cite{Telesca2016} analysed the spectral and the multifractal characteristics of several wind speed time series in Switzerland, they found cyclic components with period of $1$ day and $12$ hours. These cyclic components are linked with the daily cycle of temperature and pressure, along with persistence and multifractal characteristics at large timescales, but anti-persistence and monofractal behaviours at smaller ones.

In our paper, we apply the MFDFA to a large dataset of $119$ wind speed time series, measured by weather stations belonging to the meteorological network that covers the whole territory of Switzerland. Our aim is to investigate the spatial variability of the persistence, and the multifractal features of wind speed in Switzerland; and to find possible relationship with its topography. In order to analyse only the inner fluctuations of wind speed that are not affected by seasonal cycles, we firstly removed from the wind speed time series the trend and seasonal components and focused our attention only on the remainder (residual) time series.

\section{Data and exploratory analysis}
\label{sec:1}
\begin{figure}
%\rule{1cm}{1cm}width=\linewidth
\centering
\includegraphics[width=\linewidth]{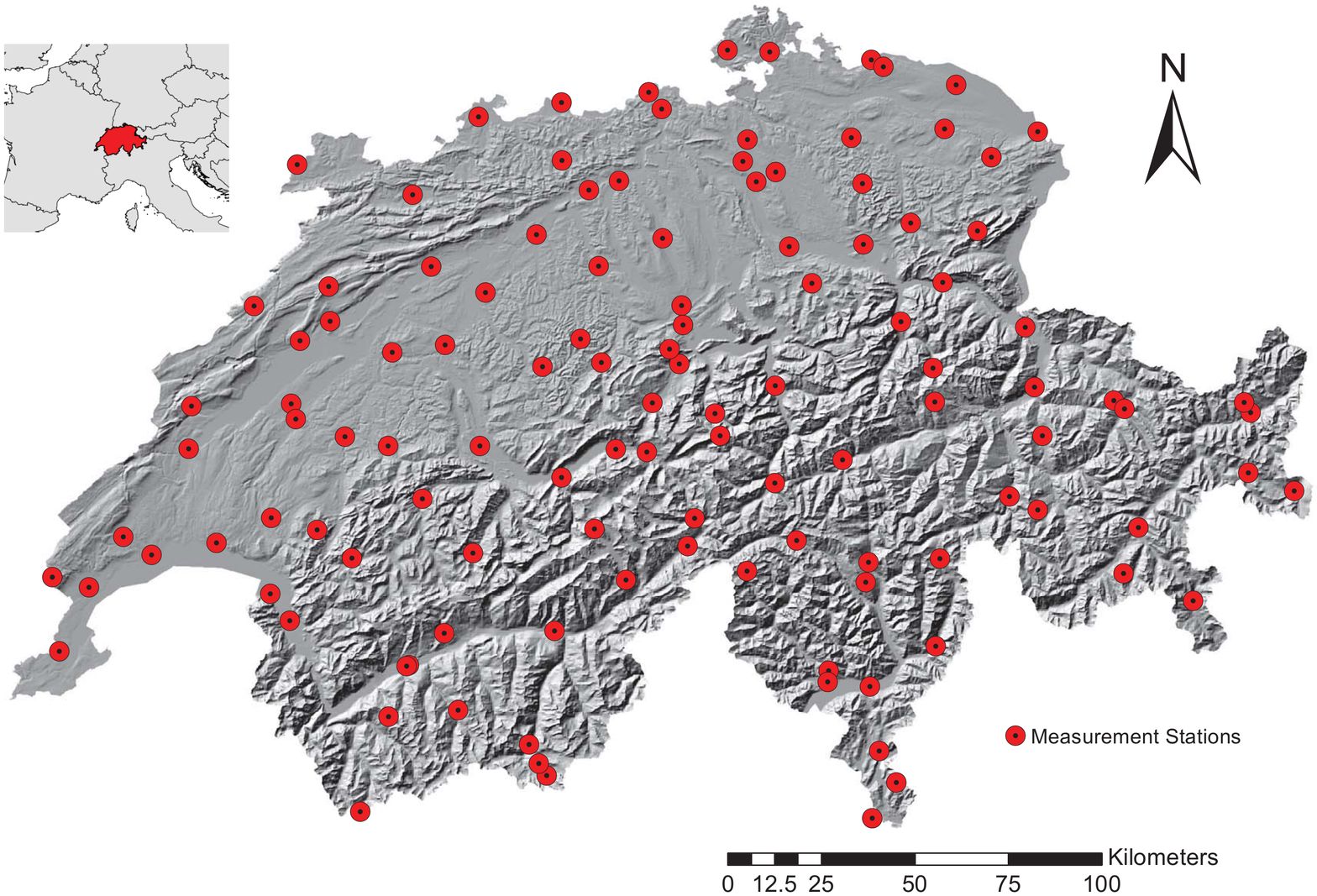}
\caption{Study area and location of wind measurement stations.}
\label{fig1}  
\end{figure}

\begin{figure}
%\rule{1cm}{1cm}width=\linewidth
\centering
\includegraphics[width=\linewidth]{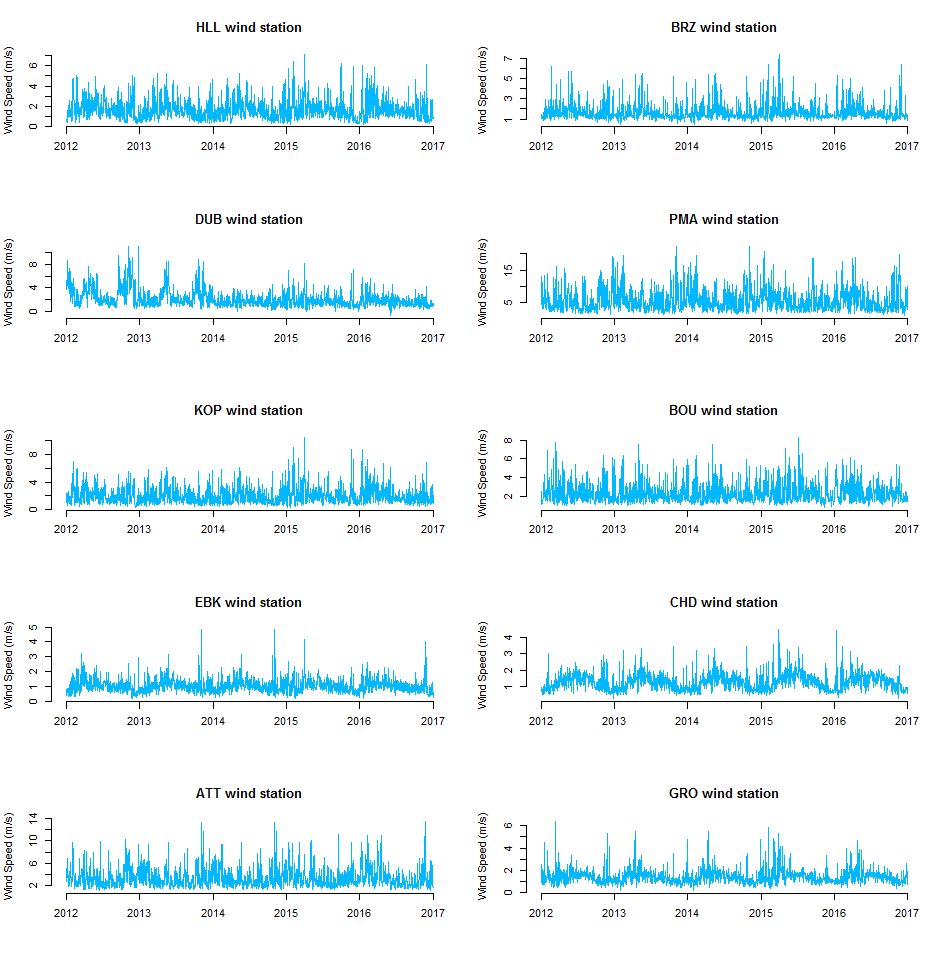}
\caption{Some wind speed time series.}
\label{fig2}  
\end{figure}

The data used in this work are provided by the Federal Office of Meteorology and climatology of Switzerland, which manages a wide network of meteorological stations covering the entire Swiss territory more or less homogeneously at different altitudes (Fig. \ref{fig1}). The raw data consist of high frequency ($10$-min sampling time) wind speed series, collected by $119$ stations, during the period between $2012$ and $2016$ (Fig. \ref{fig2} shows, as an example, some wind speed time series). We analysed the daily means of wind speed in order to remove the periodicity of one day and $12$ hours \cite{Telesca2016}.

Because of the complexity of the data, we performed an exploratory analysis in order to identify the probability distribution that better describes wind speeds. We considered the three distributions that are mostly used to model wind speed. Table $1$ illustrates for each distribution its probability density function and the corresponding parameters. 

\begin{table}
\centering 
%\footnotesize
\begin{tabular}{p{2cm}cp{4cm}}
\hline 
Distributions  &  Density function  &  Parameters \\ 
\hline  
Weibull  &  $f(x;\lambda,k)=\left\{\begin{array}{rcl}
\frac{k}{\lambda} (\frac{x}{\lambda})^{k-1} e^{-(\frac{x}{\lambda})^{k}} \qquad x\geq 0 \\
0   \qquad \qquad  \qquad x < 0
\end{array}\right.  $  &  $k$ is the shape and $\lambda$ is  the scale \cite{reisx}.  \\
\\
Gamma & $f(x;\alpha,\beta)=\frac{\beta^{\alpha}x^{\alpha-1}e^{-x\beta}}{\Gamma(\alpha)}$  & For $x \geq0$ and $\alpha,\beta > 0$ $\alpha$ is the shape and $\beta$ is the rate of the Gamma distribution \cite{qpEVT}. \\
\\

Generalised Extreme Values (GEV) & $f(x;\mu, \sigma, \xi ) = exp\{ -[1+\xi(\frac{x-\mu}{\sigma})]^{\frac{-1}{\xi}} \} $ & where $ 1+ \xi(x-\mu)/\sigma > 0$, and $\mu $ is the location parameter, $\sigma $ is the scale parameter and $\xi$ is the shape \cite{colesB}.\\

\hline

\end{tabular}
\caption{Description of the used probability distributions.}
\label{somenoise}
\end{table} 

In order to evaluate the goodness-of-fit of the data with each probability distributions, we used the well-known Kullback-Leibler divergence (KL). Given a random sample $X_1, \ldots, X_n$ from a probability distribution $P(x)$ with density function $p(x)$ over a non-negative support. If we suppose that the sample comes from a specific probability distribution $Q(x)$ with a density function $q(x)$, the KL information on the divergence between $P(x)$ and $Q(x)$  is given by the following formula \cite{kullback1951}:

\begin{equation}
D_{KL}(p\|q)=\int_0^\infty p(x) ln \frac{p(x)}{q(x)}dx.
\end{equation}

It is known that the information divergence $D_{KL}(p\|q) \geqslant 0$. Therefore, if $D_{KL}(p\|q)=0$, the sample comes from the specific probability distribution $Q(x)$ \cite{Ebrahimi1992, waal1996}.

\begin{figure}
%\rule{1cm}{1cm}width=\linewidth
\centering
\includegraphics[scale=0.5]{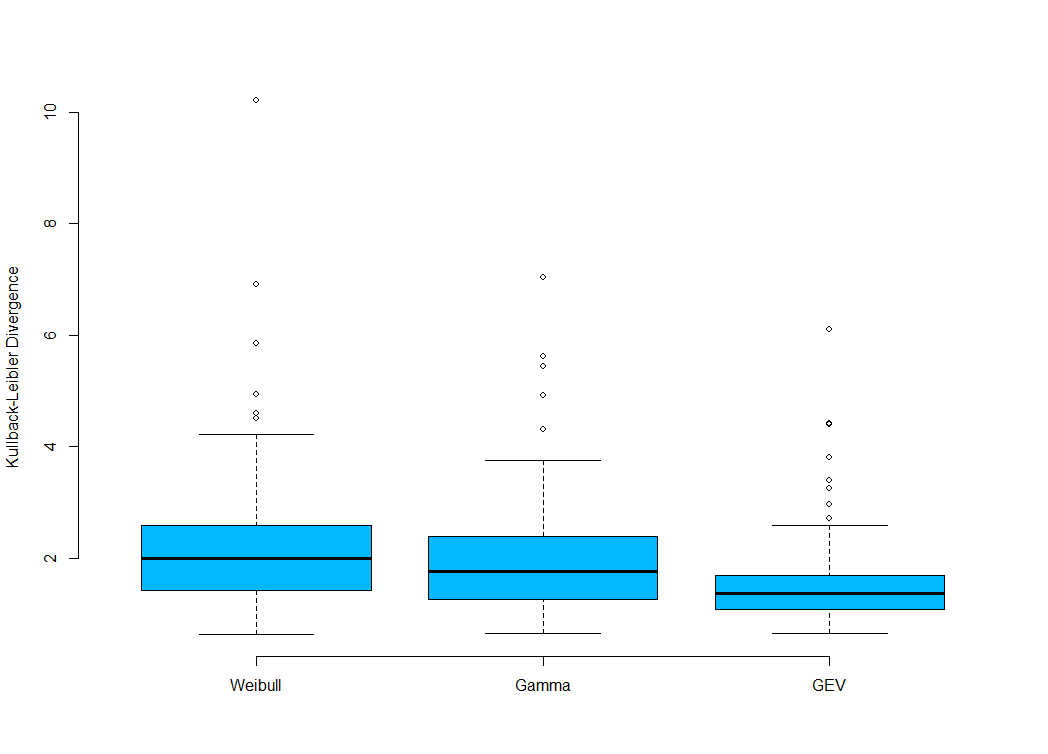}
\caption{Values of the Kullback-Leibler Divergence for each probability distribution (each value corresponds to a different measurement station).}
\label{fig3}  
\end{figure}

Therefore, we calculated the KL divergence between the proposed probability distributions and the wind speed data. We obtained the results shown in Fig. \ref{fig3}, which suggests that the GEV distribution fits the data better than the other two distributions.

Before applying the MFDFA, we decomposed the time series by using the Seasonal and Trend decomposition based on the Loess smoother (STL), proposed by \cite{Cleveland1990}. In this method each wind speed time series is decomposed into trend ($T_i$), seasonal ($S_i$) and remainder ($R_i$ ) components.

The STL decomposition consists mainly of two important recursive procedures: an inner loop and an outer loop. The inner loop is used to update the trend and seasonal components, while the outer loop computes the robustness weights according to the remainder component, which will be used in the next iteration of the inner loop. The outer loop tends to reduce the weights of outliers or extreme values in the time series; in other words, the weights decrease by increasing the distance from $x_i$ whereas the closest point to $x_i$ has the largest weight. For more details on the STL decomposition, the reader can refer to Cleveland et al. \cite{Cleveland1990}.

\begin{figure}
%\rule{1cm}{1cm}width=\linewidth
\centering
\includegraphics[scale=0.5]{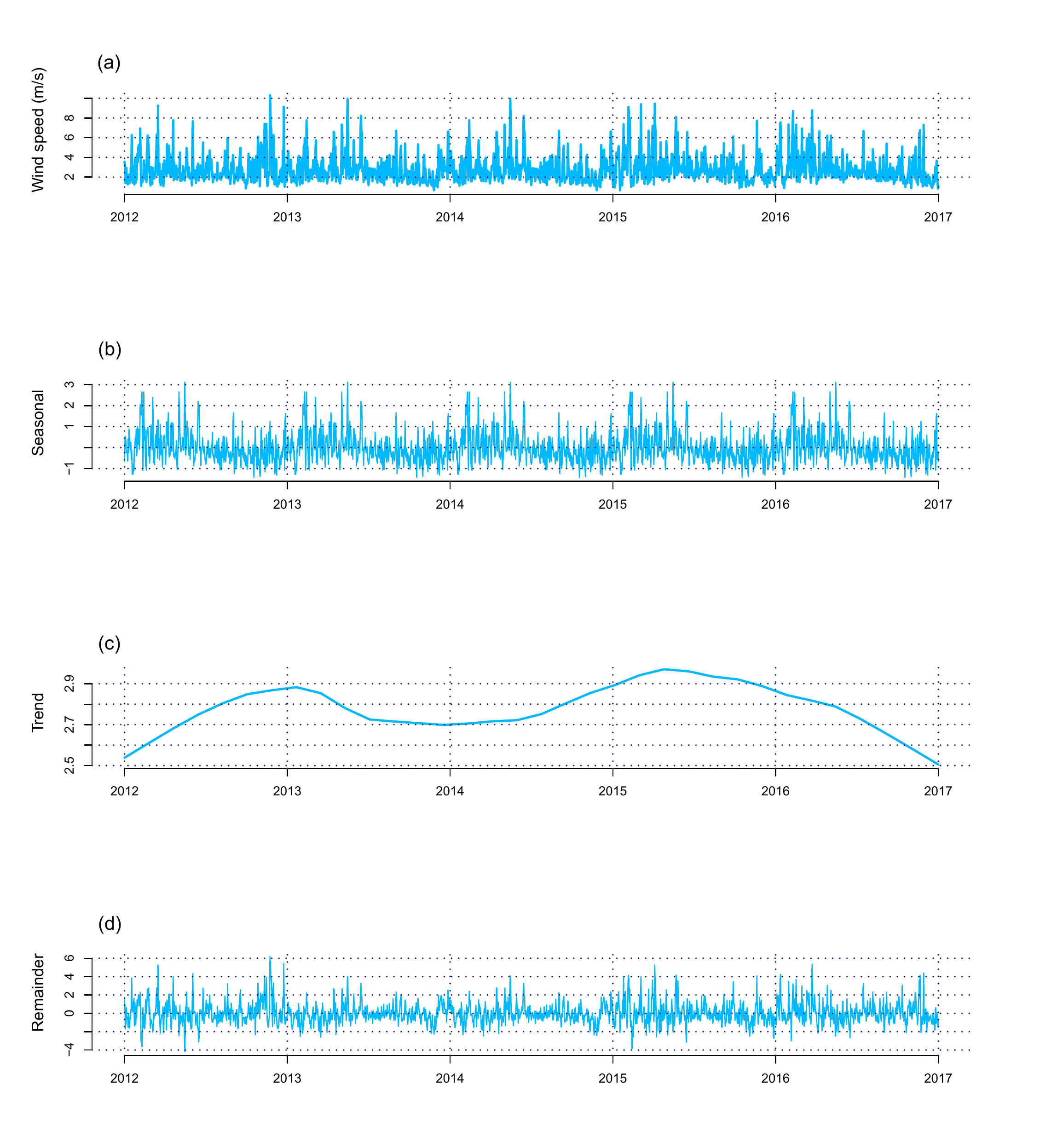}
\caption{STL decomposition of wind speed time series (Jungfraujoch station: $7^{\circ}59$/$46^{\circ}33$ of longitude/latitude) $3580m$ of altitude.}
\label{fig4}  
\end{figure}

Fig. \ref{fig4} shows, as an example, the three components for the station Jungfraujoch: the seasonal (Fig \ref{fig4}.b), the trend (Fig \ref{fig4}.c), and the remainder (Fig \ref{fig4}.d). As it can be clearly seen, the seasonal component is characterized by the annual oscillation, which is linked with the yearly meteo-climatic cycle. The trend component shows a slow time evolution of the wind speed, characterized by a very small range of variability. The remainder, instead, appears quite irregular and characterized by high frequency fluctuations, suggesting a "richer" dynamics that could be probably linked with the local topographic conditions of the measuring site. The apparent noisy character of the remainder component does not mean that this is a purely random noise; in fact, the application of the MFDFA to the remainder component will help to capture the dynamical features of the inner fluctuations of the wind speeds.

\section{Multifractal detrended fluctuation analysis}

The most powerful technique to detect multifractality in a time series is the multifractal detrended fluctuation analysis (MFDFA) \cite{ KANTELHARDT2002}. Let $x(i)$ for $i=1,\ldots,N$ be a possibly non-stationary time series, where $N$ indicates its length. We, first, construct the "trajectory" or "profile" by integration after subtracting from the time series its average $x_{ave}$ 

\begin{equation}
Y(i)=\sum_{k=1}^i[x(k)-x_{ave}].
\end{equation}

The profile is sub-divided into $N_s=int(N/s)$ non-overlapping windows of equal length $s$. Since the length $N$ of the series may not be an integer multiple of the window size $s$, and a short part of the profile $Y(i)$ at the end may be disregarded by the procedure, the sub-division is performed also starting from the opposite end, obtaining a total of $2N_S$ segments. A polynomial of degree $m$ fits the profile in each of the $2N_S$ windows and the variance is calculated by using the following formula:
\begin{equation}
F^2 (s,v)=\frac{1}{s} \sum_{i=1}^s\{Y[(v-1)s+i]-y_v (i)\}^2.
\end{equation}

For each segment $v$, $v=1,\ldots,N_s$  and
\begin{equation}
F^2 (s,v)=\frac{1}{s} \sum_{i=1}^s\{Y[N-(v-N_s )s+i]-y_v (i)\}^2.
\end{equation}

For $v=N_{(s+1)},\ldots,2N_s$. Here $y_v (i)$  is the fitting polynomial in segment $v$. Then, averaging over all segments the $q_{th}$ order fluctuation function is computed

\begin{equation}
F_q (s)=\left\lbrace \frac{1}{2N_s} \sum_{v=1}^{2N_s}[F^2 (s,v)]^{\frac{q}{2}}\right\rbrace ^{1/q}  
\end{equation}
where, in general, the index variable $q$ can take any real value except zero. The parameter $q$ enhances the small fluctuations if negative, otherwise, the large ones if positive. $F_q (s)$ will increase with increasing $s$, and if $F_q (s)$  follows a power-law, the series exhibits a linear behaviour in the log-log scale for that specific $q$. In this case

\begin{equation}
F_q (s) \propto s^{h_q}.
\end{equation}
The exponent $h_q$ is called generalized Hurst exponent due to the equivalence between $h_2$  and the Hurst exponent $H$ \cite{Jens1988} for stationary series, leading to consider the well-known detrended fluctuation analysis (DFA) \cite{PENG1995} a particular case of the MFDFA for $q=2$. For $q=0$ the value $h_0$ corresponds to the limit $h_q$ for q → 0, and is obtained through the logarithmic averaging procedure:

\begin{equation}
F_0(s) \equiv exp \left\lbrace \frac{1}{4N_s} \sum_{v=1}^{2N_s} ln[F^2 (s,v)]\right\rbrace  \propto s^{h_0}.
\end{equation}

In general, the exponent $h_q$ will depend on $q$, and it monotonically decreases with the increase in $q$, the series is multifractal. If $h_q$ does not depend on $q$ the series is mono-fractal.
Multifractal series can be also studied by means of the singularity spectrum, obtained by applying the Legendre transform \cite{Parisi1983}. From the relationship

\begin{equation}
\tau(q)=qh_q-1
\end{equation}
and 
\begin{equation}
\alpha=d\tau/dq
\end{equation}
we obtain
\begin{equation}
f(\alpha)=q\alpha-\tau(q)
\end{equation}
where $\alpha$ is the Hölder exponent and $f(\alpha)$ indicates the dimension of the subset of the series that is characterized by $\alpha$ \cite{Ashkenazy2003}. The multifractal spectrum indicates how much dominant are the various fractal exponents present in the series. The width of the singularity spectrum as well as the range of the generalized Hurst exponent $(max(h_q)-min (h_q))$ are often used to quantitatively measure the degree of multifractality of the series, thus the wider the spectrum more multifractal the series.

\section{Results}
We applied the MFDFA to the remainder components of the wind speed time series recorded at each measurement station. As we pointed out in section $2$, the remainder components do not present the seasonal oscillations that would be originated by the annual and sub-annual meteo-climatic cycles. Applying the STL decomposition to remove the seasonal components does not only produce more reliable results but is likely to give information on the inner dynamics of wind speeds.

\begin{figure}
%\rule{1cm}{1cm}width=\linewidth
\centering
\includegraphics[width=\linewidth]{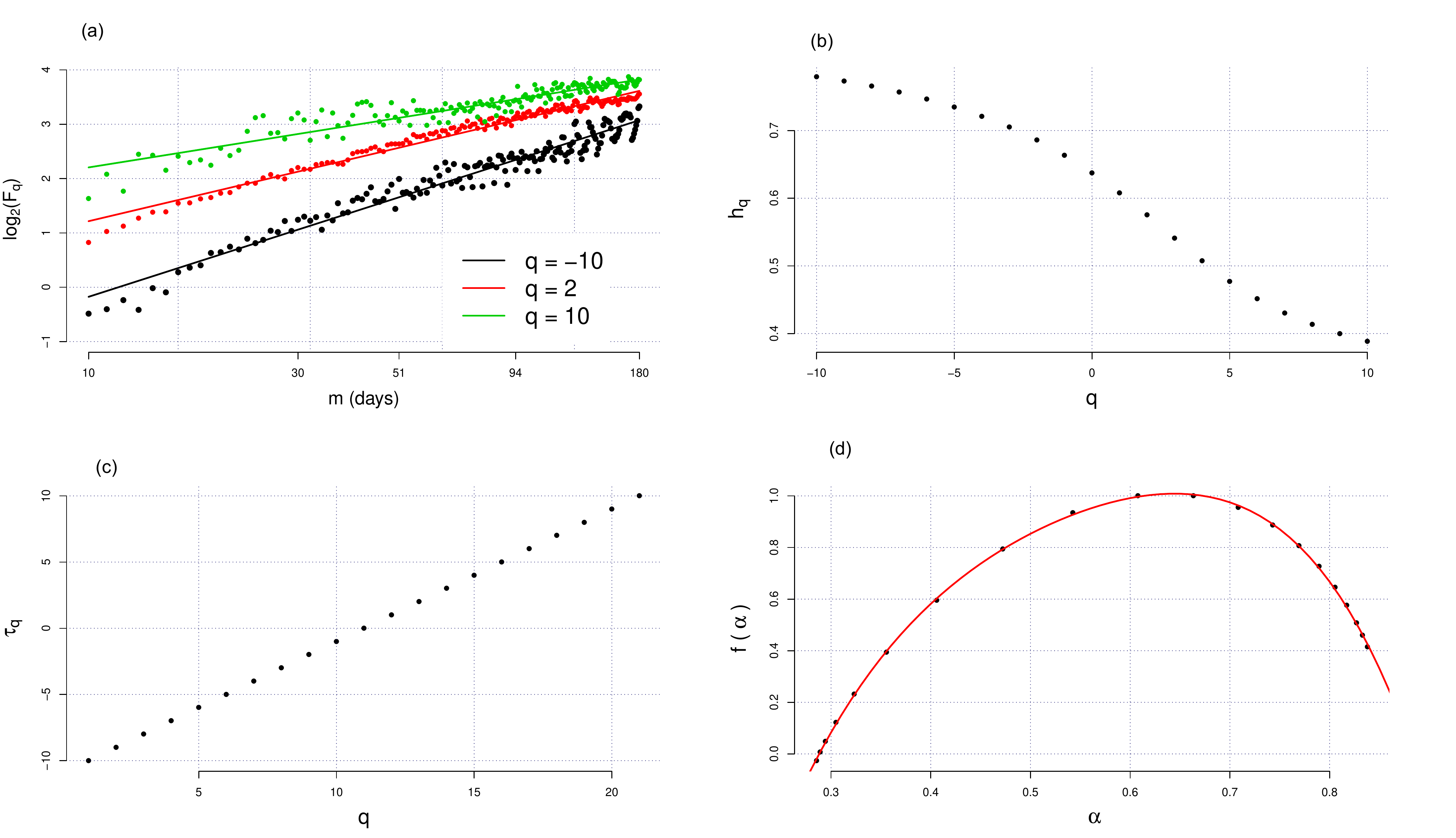}
\caption{The MFDFA results of wind speed time series from the Jungfraujoch measuring station.}
\label{fig5}  
\end{figure}

Fig. \ref{fig5} shows, as an example, the results of MFDFA applied to the remainder component of station Jungfraujoch (shown in Fig. \ref{fig4}d). We chose the second-degree (i.e. $m=2$) detrending polynomial to perform the MFDFA \cite{Telesca2016} and the time scales range from $10$ days to $180$ days (the upper time scale is constrained by the size of time series, which is $5$ years long). As it can be clearly seen, the fluctuation functions are well fitted by a straight line in log-log scales (Fig. \ref{fig5}a), and this indicates the presence of scaling at any $q$. In particular, $F_2$ allows us to estimate the Hurst exponent that, for stationary series, is given by the scaling exponent calculated for $q=2$. In this case, $H = 0.570 \pm 0.006$, suggesting that the remainder component is characterized by a weak persistence. Fig. \ref{fig5}b shows the dependence of the generalized Hurst exponent with $q$: we see that it is a decreasing function and this indicates the presence of multifractal features in the time fluctuations of the remainder. Fig. \ref{fig5}c shows the function $\tau_q$ and Fig. \ref{fig5}d the multifractal spectrum obtained by the Legendre transform. The multifractal spectrum shows the well-known single-humped shape that characterizes multifractal series.  
In order to quantify the multifractal characteristics of the series, we fitted the multifractal spectrum by a $4^{th}$ order polynomial and calculated $(1)$ the width as $W=(\alpha_2- \alpha_1)$, where $\alpha_2 > \alpha_1$ are the two zero-crossings of the fitting function, and $(2)$ the asymmetry as the ratio $A=(\alpha_0-\alpha_1)/(\alpha_2-\alpha_0)$, where $\alpha_0$ is the maximum of the multifractal spectrum. 

The width $W$ is a quantitative measure of the multifractal degree, thus the larger the width the more multifractal the series. The asymmetry $A$ measures the skewness of the multifractal spectrum. So if $A=1$, the multifractal is symmetric, which indicates that  the low and high fluctuations govern more or less equally the dynamics of the series. If $A>1$, the multifractal spectrum is left-skewed (indicating a dominance of the large fluctuations), while if $A<1$, the multifractal spectrum is right-skewed (indicating a relative dominance of the small fluctuations). For the multifractal spectrum shown in Fig. \ref{fig5}, we found $W=0.55$ and $A=2.16$; thus the remainder of the wind speed measured at Jungfraujoch station is rather multifractal with the time dynamics mainly governed by the large fluctuations.

\begin{figure}
%\rule{1cm}{1cm}width=\linewidth
\centering
\includegraphics[width=.9\linewidth]{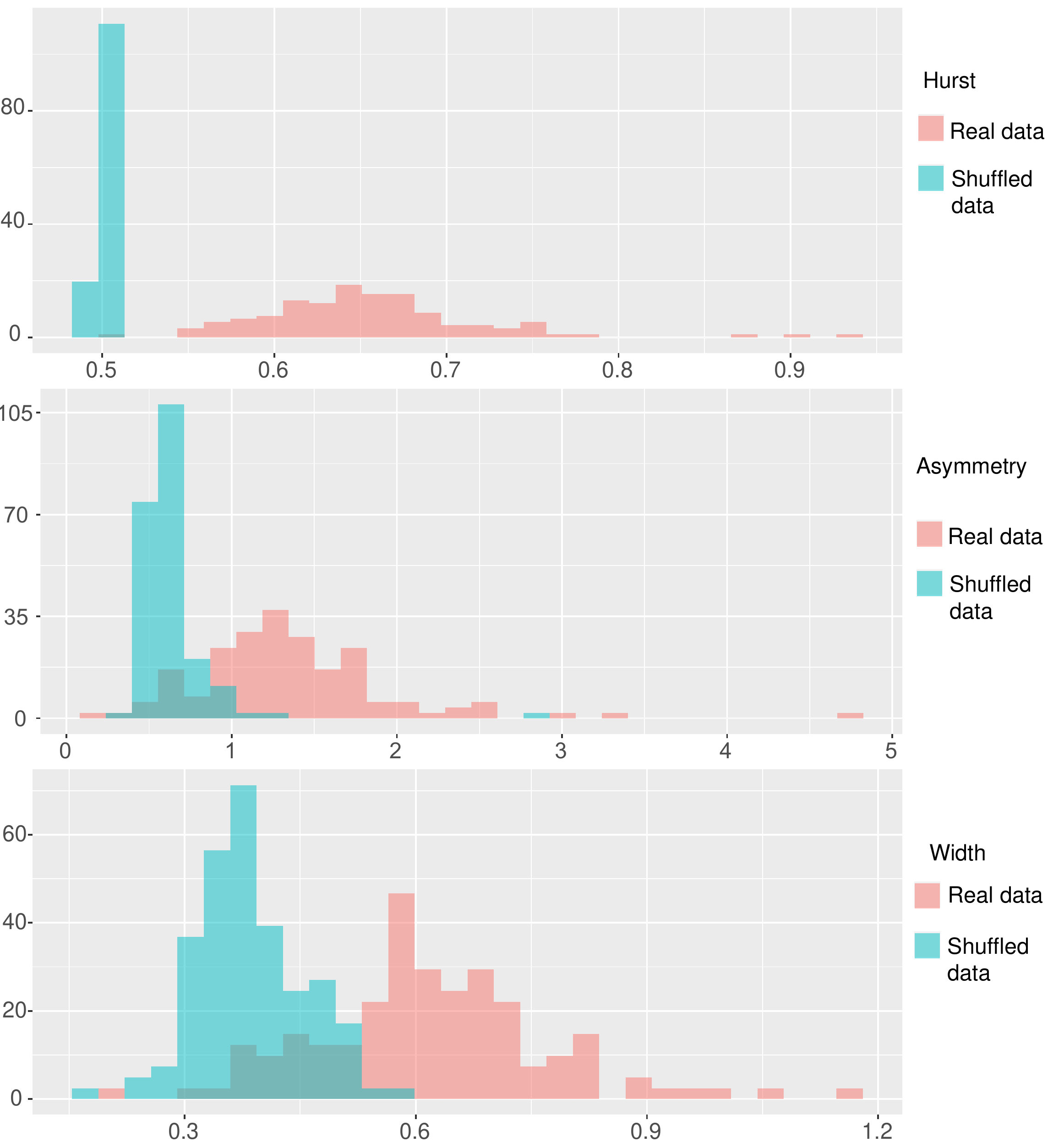}
\caption{histogram of the MFDFA parameters: original data versus shuffled data.}
\label{fig6}  
\end{figure}

We applied the MFDFA to the remainder components and we calculate the MFDFA parameters ($H$, $W$, $A$)
of all the $119$ wind speed time series. Fig. \ref{fig6} shows the distributions of these three parameters: $H$ ranges between $0.51$, and $0.94$, $W$ between $0.206$ and $1.15$ and $A$ between $0.19$ and $4.77$. 
In order to check if the results for the three parameters are due to chance, we generated $1,000$ shuffled data for each time series and calculated $H$, $W$ and $A$; then for each station, we averaged the results. Fig. \ref{fig6} shows also the distribution of the means of $H$, $W$ and $A$ calculated for the shuffled data. We can clearly see that the distribution of the mean H for the shuffled data is very narrow and peaked at $0.5$, which is the Hurst exponent of a purely random series; the shuffling, in fact, is a randomisation that preserves the global distribution of the data and destroys the temporal patterns. Thus, comparing the distribution of the $H$ values of the original series with that of the shuffled data, we can argue that the original series are significantly persistent, since the distribution of their $H$ values is different from that of the mean $H$ values calculated for the shuffled data.

\begin{figure}
%\rule{1cm}{1cm}width=\linewidth
\centering
\includegraphics[width=.95\linewidth]{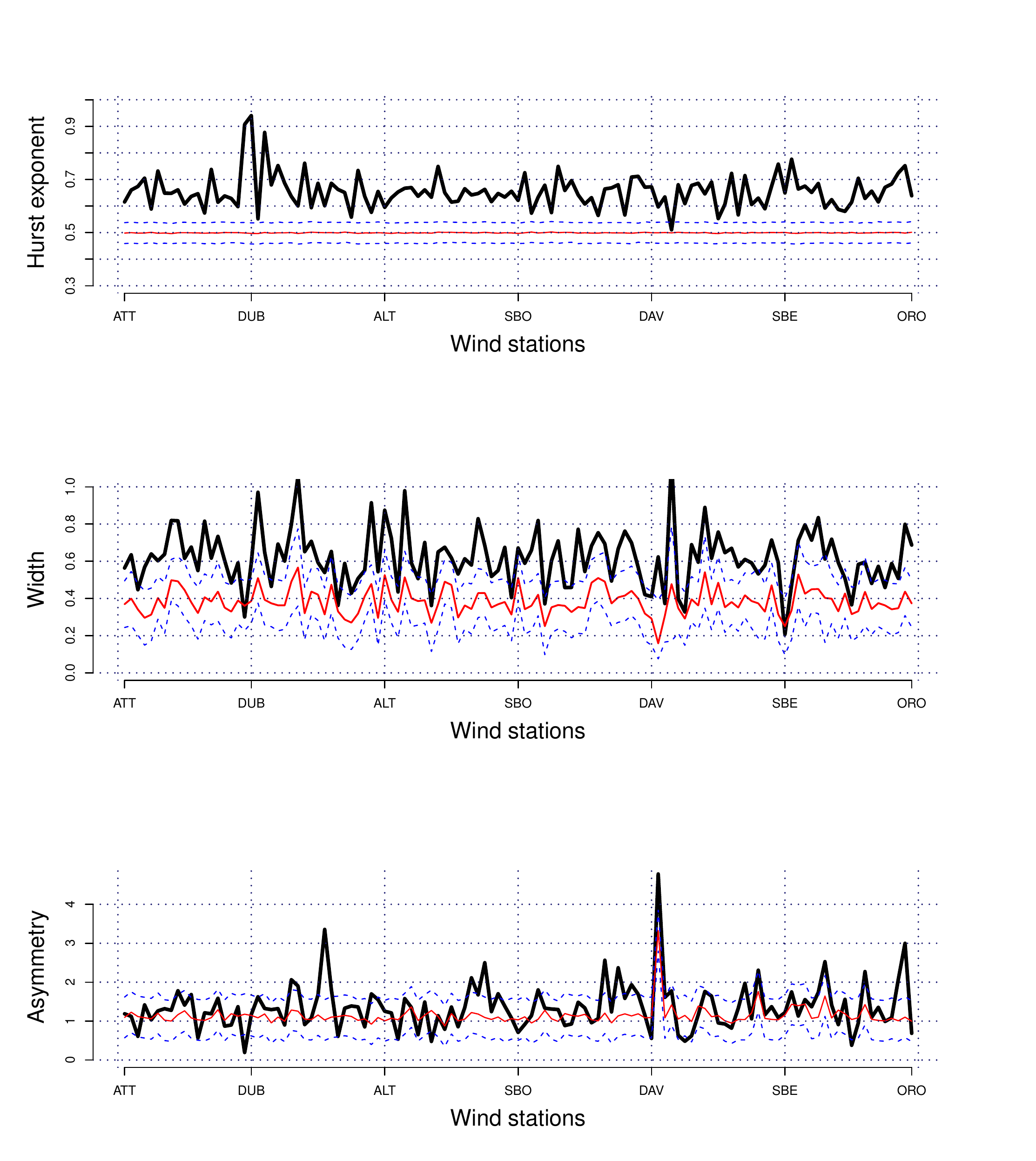}
\caption{Variation of $H$, $W$, and $A$ of the original time series (black) with the mean of shuffled time series (red) $\pm$ standard deviation (blue).}
\label{fig7}  
\end{figure}

Fig. \ref{fig7} shows the variation of $H$, $W$ and $A$ for the original series versus the station, along with the mean $H$, mean $W$ and mean $A$ ($\pm$standard deviation) for the shuffled data: the results confirm what we have observed from the comparison between the distributions of the multifractal parameters of the original series with those of the shuffled data.

We mapped the spatial distribution of the multifractal parameters
($H$, $W$ and $A$) of the original series on the territory of Switzerland, to see if any link with the topography of the area could be detected. To this aim, we used a well-known machine learning algorithm, called Extreme Learning Machine (ELM) \cite{ELM}. 

\begin{figure}
\begin{center}
\includegraphics[width=.7\linewidth]{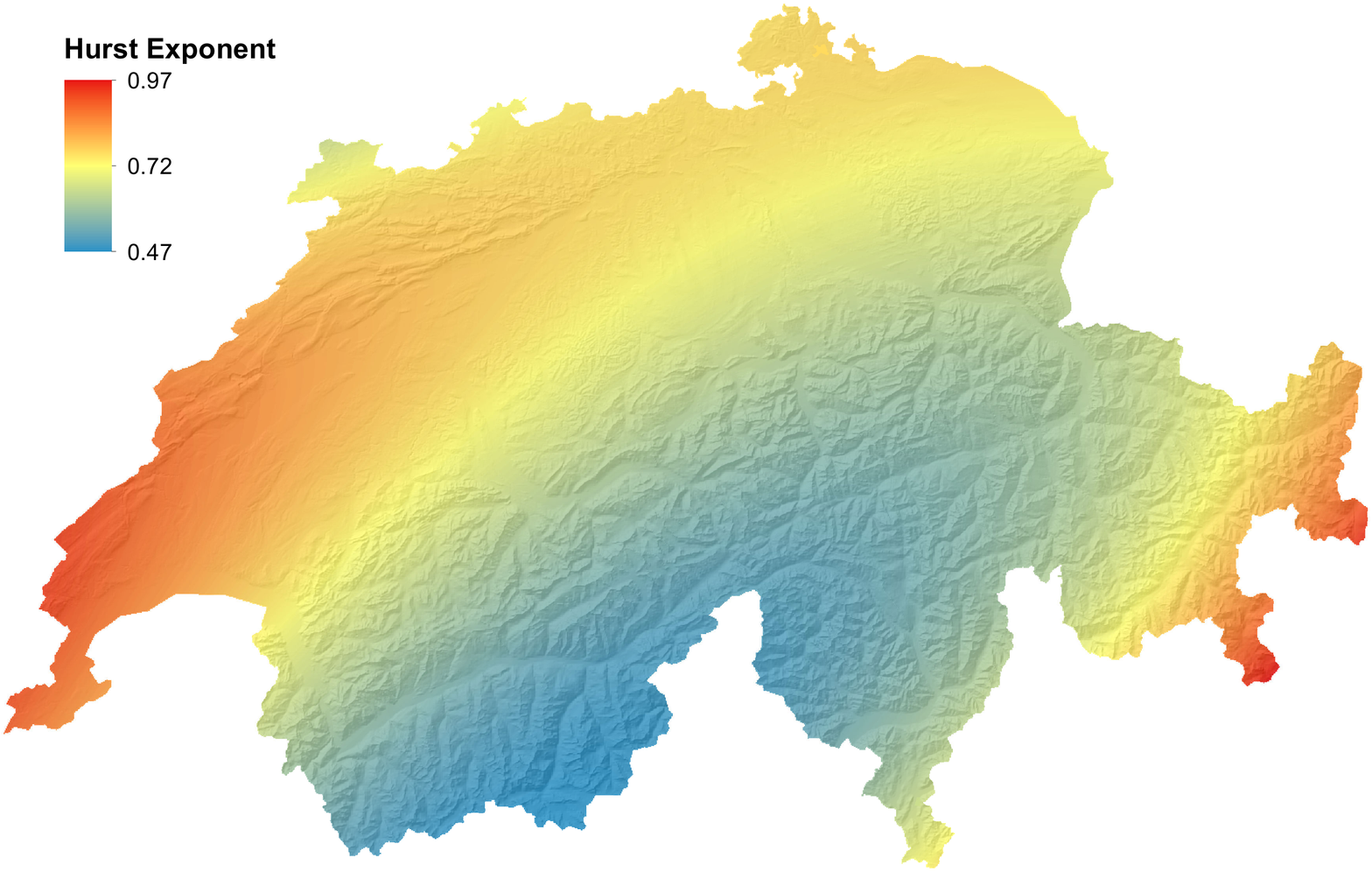}
\includegraphics[width=.7\linewidth]{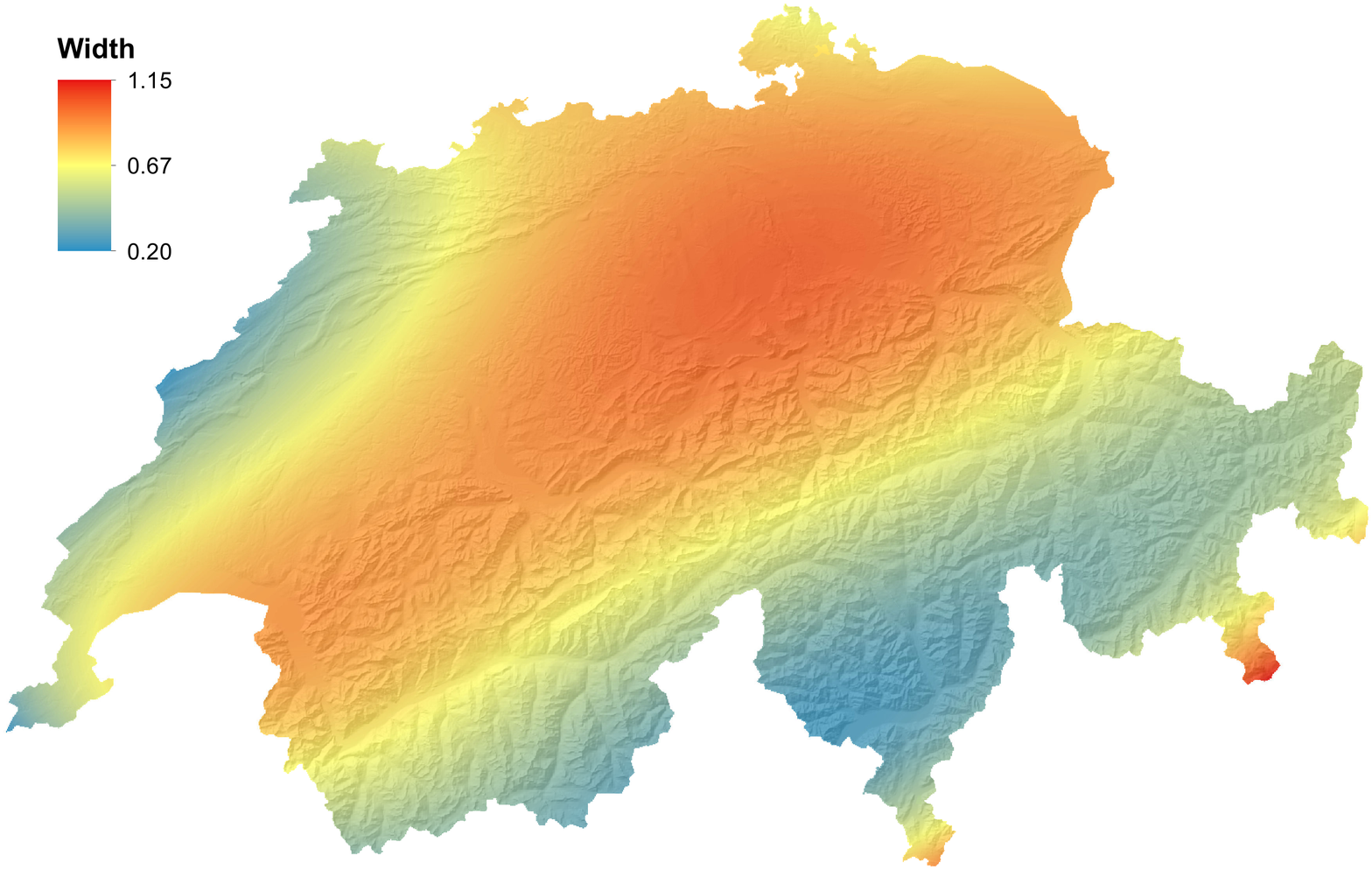}
\includegraphics[width=.7\linewidth]{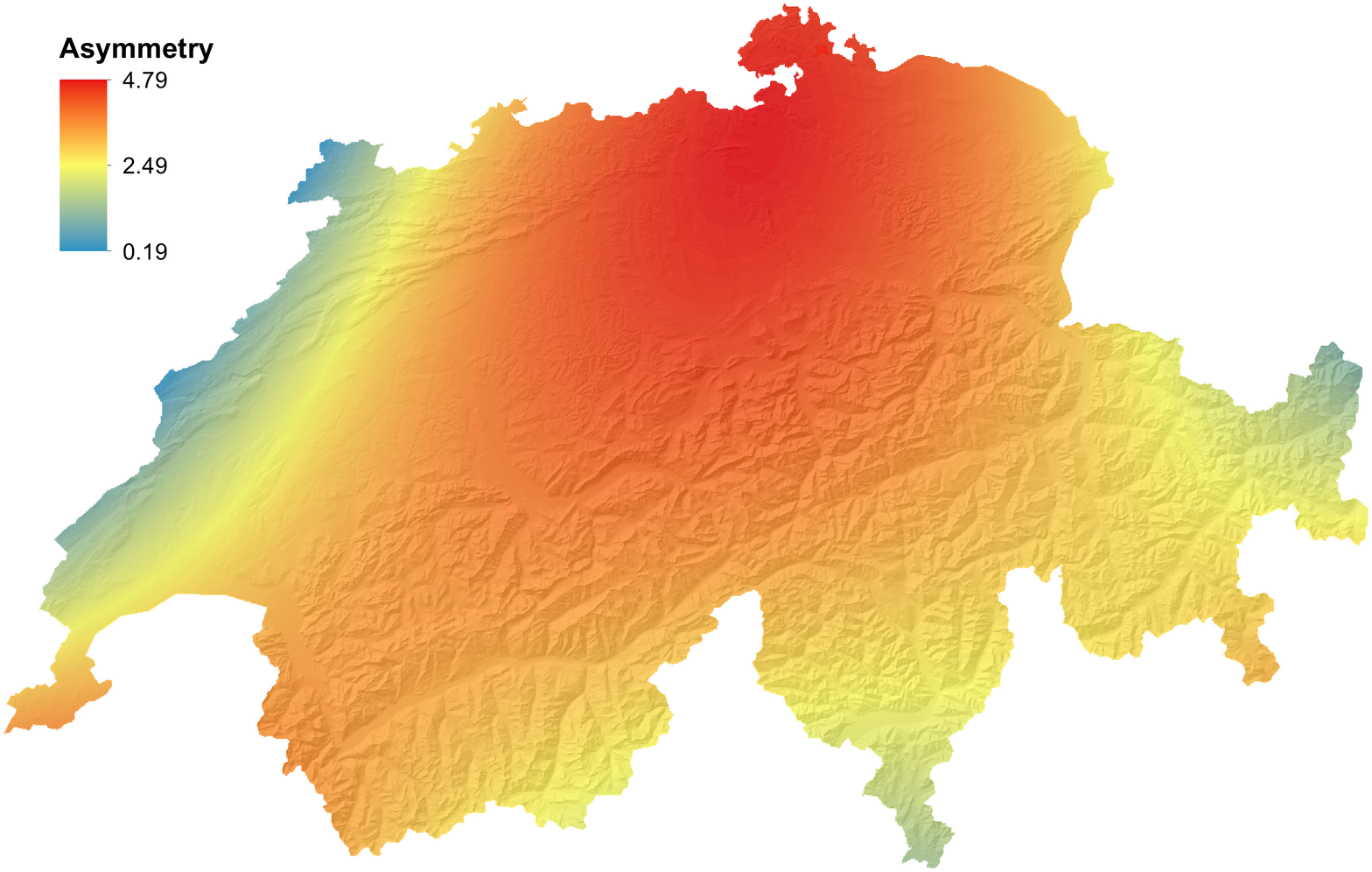}
\end{center}
\caption{Spatial mapping of MFDFA parameters of the original data.}
\label{fig8}  
\end{figure}

\begin{figure}
\begin{center}
\includegraphics[width=.7\linewidth]{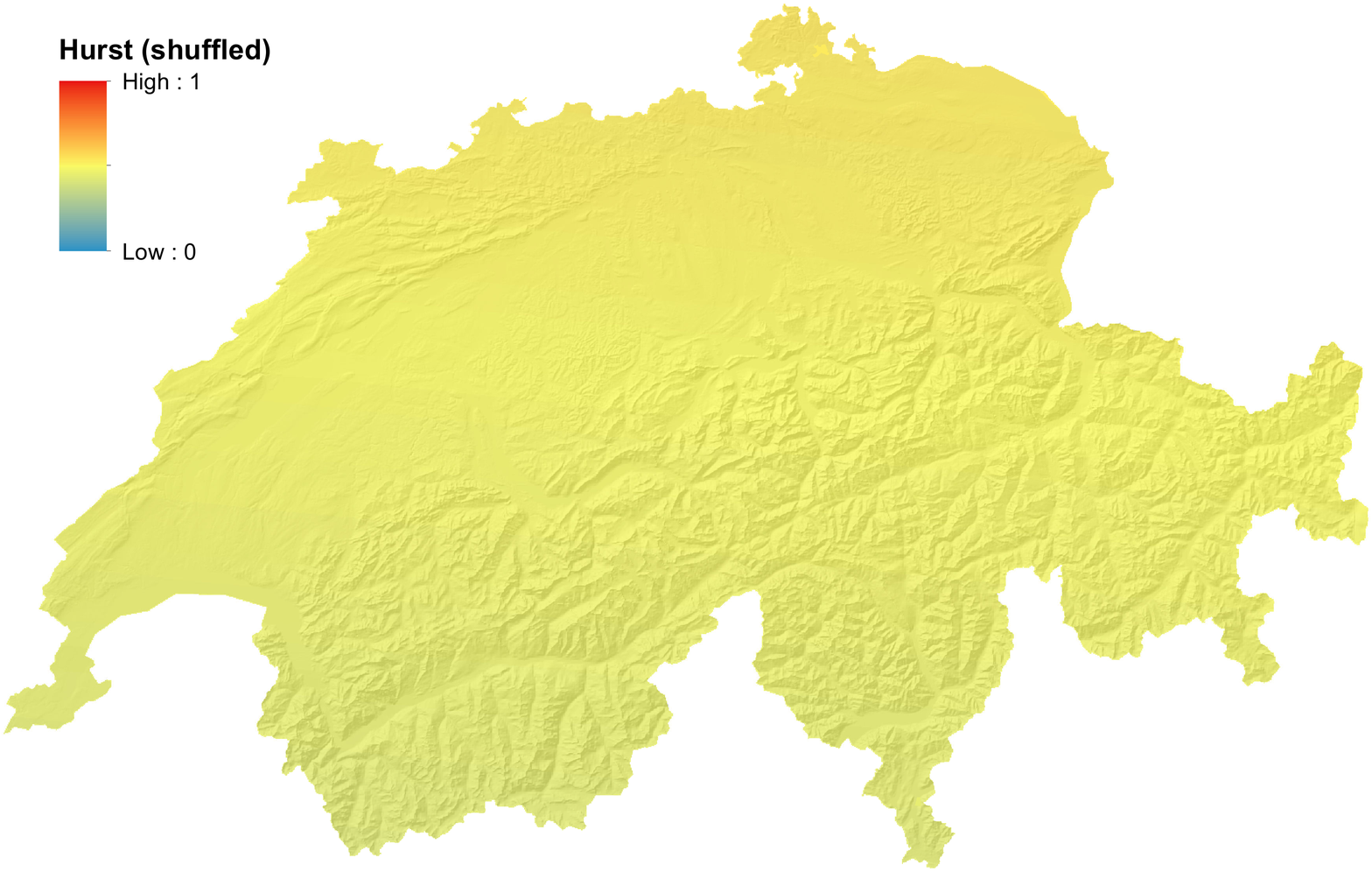}
\includegraphics[width=.7\linewidth]{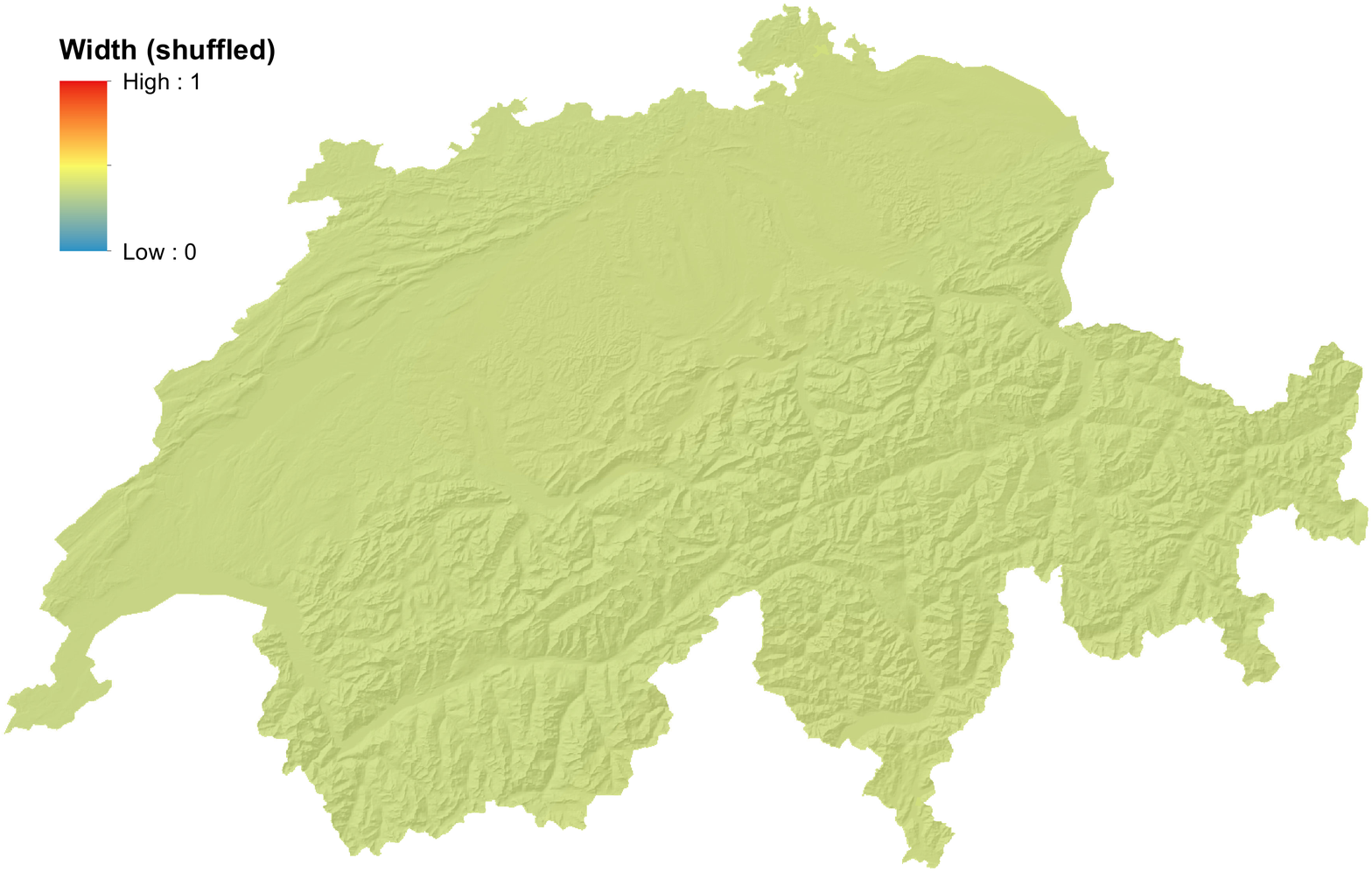}
\includegraphics[width=.7\linewidth]{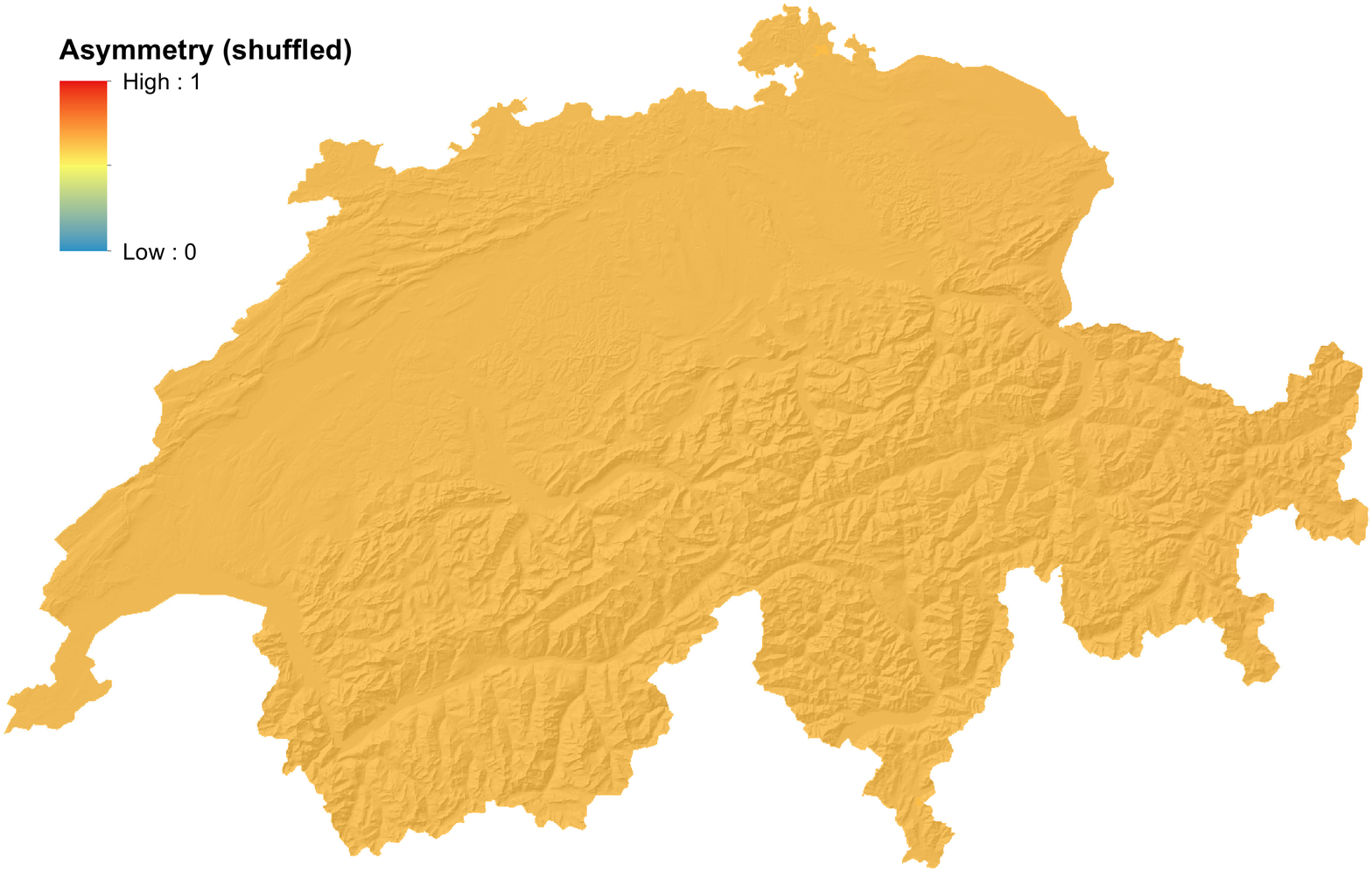}
\end{center}
\caption{Spatial mapping of MFDFA parameters of the shuffled data.}
\label{fig9}  
\end{figure}

Fig. \ref{fig8} shows the results obtained by using ELM, which learned the existing structure between the input (i.e. the XY-coordinates) and the output (i.e. each MFDFA parameter). The resulting maps are obtained by using a $2$-dimensional grid of $250m$ of resolution. In order to evaluate whether or not the obtained spatial structure was due to chance, we mapped the spatial distribution of the MFDFA parameters ($H$, $W$ and $A$) of the shuffled data.  Fig. \ref{fig9} shows the maps  for the shuffled data. We see that there is no more structure, which means that the results of the original data are not due to chance. 

\section{Conclusion}
The daily means of wind speed at $119$ different weather-monitoring stations in Switzerland were analysed in terms of their multifractal properties. A preliminary filtering of the data was performed by using the STL algorithm that decomposes a time series into trend, seasonal and remainder components. Our study was focused on the multifractal analysis of the remainder component of each wind speed time series, which could convey information about the inner time fluctuations of the wind speed, likely linked with the local topographic conditions of the measuring site. 

All the analysed remainders of wind speed are significantly persistent, being characterized by a Hurst exponent significantly larger than $0.5$. This suggests that the temporal fluctuations of wind speed are characterized by a relative dominance of low frequency fluctuations with respect to the high frequency ones. Persistence implies that the process is generated by positive feedback mechanisms, as it can be found in climatic processes (e.g. Mandelbrot and Wallis, $1969$) or in the statistics of ocean waves \cite{Jens1988} where it is possible to observe rather clear trends in the wave-height with relatively little noise \cite{Cuomo2000}.

The multifractality of the series was quantified through the width and the asymmetry of the multifractal spectrum.  Both parameters show a certain variability among the sites, which could be linked with the typical morphology and complex topography of the Swiss territory. The Swiss territory can be sub-categorized into mountain ($\geq 1200 m$ a.s.l.), valley (in Alpine terrain), and rolling to flat terrain (the Swiss Plateau), which comprises the region situated between the Jura range and the Alps, and a number of lower-elevation locations in the Jura range and one low-land location south of the Alps \cite{Stucki2016}. 

The spatial distribution of the $W$ and $A$ shows nearly clearly the distinction between these three topographic categories; both parameters assume the highest range of values in the Swiss plateau, indicating that the wind speed measured in this area are characterized by a larger dynamical heterogeneity and a higher dominance of the relatively larger fluctuations. The higher heterogeneity and asymmetry of the wind speed in the Swiss plateau is probably due to the typical morphology of the area, in which the distance between the Alps and the Jura Mountains becomes narrower from east to west, and is quite small in the region of Lake Geneva. The air blowing from the north-east is canalised between these two mountain ranges, producing an increase of wind speed towards western Switzerland. It is also likely that such "canalizing" effect produces some local air turbulence phenomena, which could lead to intermittent wind speed fluctuations and could be responsible of the higher multifractality of the wind speed series measured in this area.

The present study, of course, cannot exhaustively explain the very complex spatio-temporal variability of wind speed over the entire Swiss territory; however, it represents a contribution to the characterization of wind speed and to the understanding of the mechanisms that govern its dynamics in a territory, like that of Switzerland, where several advective weather situations occur.  

\section{Acknowledgements}
The authors thank MeteoSwiss for giving access to the data via IDAWEB server. LT thanks the support of Herbette Foundation. This research was partly supported by the Swiss Government Excellence Scholarships.
%\nocite{*}

\bibliography{xampl}
\bibliographystyle{elsarticle-num}

\section{Appendix}
\subsection{Extreme Learning Machine}
\begin{figure}
%\rule{1cm}{1cm}width=\linewidth
\centering
\includegraphics[width=\linewidth]{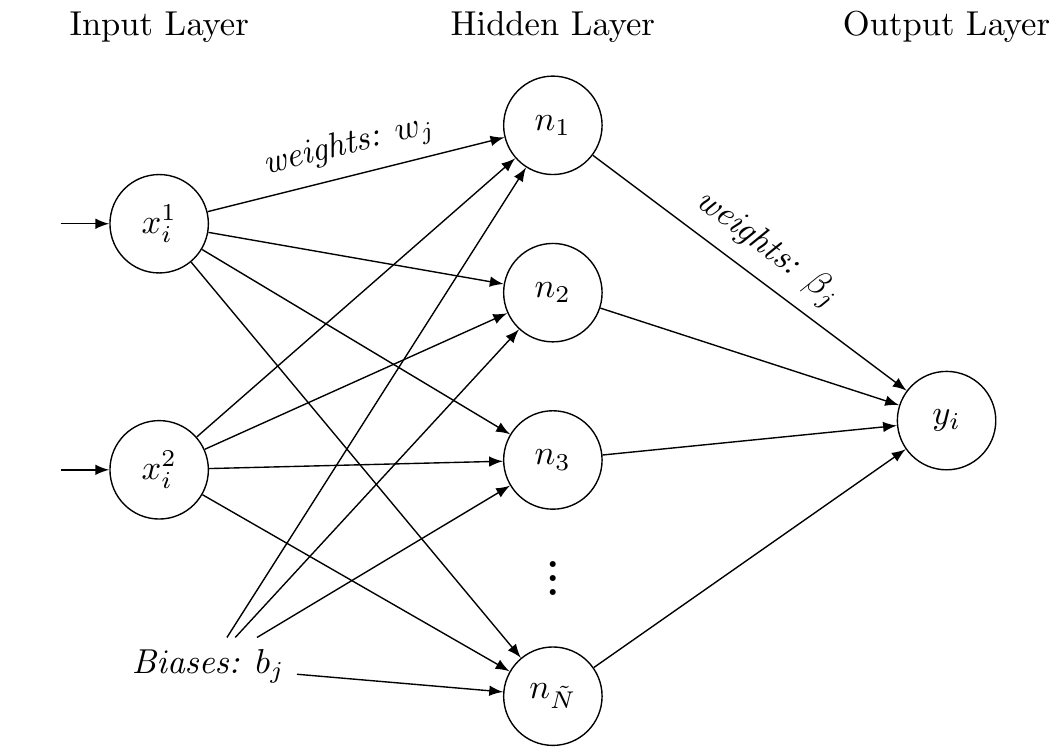}
\caption{Extreme Learning Machine architecture used in this study (adapted from \cite{Leuenberger_Dec15}).}
\label{figx}  
\end{figure}

Extreme Learning Machine (ELM) is a single-hidden layer feed-forward neural network (Fig. \ref{figx}). Let $(x_i, y_i)_{i=1, \ldots, n}$ be $n$ data points, where $(x_i^1, x_i^2, \ldots, x_i^d)^T \in \mathbb{R}^d$ and $y_i \in \mathbb{R}$. 

For a fixed number of hidden nodes $\widehat{N}$, ELM generates random weights $w_j$ $(j=1, \ldots, \widehat{N})$ that connect the input layer and the hidden layer, and the biases $b_j$ for each node. Then, it computes the $n \times \widehat{N}$ matrix consisting of the outputs of the hidden layer, which is given by the following formula:
\begin{equation}
H_{ij}=g(x_{i}.w_{j}+b_{j})
\end{equation}
where $g$ is an infinitely differentiable activation function \citep{ELM, Leuenberger_Dec15, GOLAY2017}.

To obtain the weights $\beta_j$ $(j=1, \ldots, \widehat{N})$ between the hidden layer and the output layer, ELM uses the Moore-Penrose generalized inverse of the matrix $H$:

$\beta=H^{\dagger} y$.

Once all the weights and biases are defined, we used a testing dataset consisting of $20\%$ of the original data to estimate the generalisation error. The choice of ELM is motivated by the fact that it is fast (with only one parameter to optimise, which is the number of hidden nodes $\hat{N}$), gives accurate predictions, and it was proven to be a universal modelling algorithm \citep{ELM}. The computations of this study are performed by using the elmNN library \cite{elmp} of the R software \cite{lanR}.

\newpage

% Appendix

\end{document}